\documentclass[apj,numberedappendix]{emulateapj}
\usepackage{apjfonts}
\def\be{\begin{equation}}
\def\ee{\end{equation}}
\def\bea{\begin{eqnarray}}
\def\eea{\end{eqnarray}}
\def\nn{\nonumber}
\def\bt{\vec{\theta}}
\def\cf{\mathcal{F}}
\def\bcf{\vec{\cf}}
\def\ch{\mathcal{H}}
\def\cw{W_{\rm len}}
\shorttitle{High Convergence Regions}
\shortauthors{Wang, Haiman, and May}
\begin{document}
\title{Constraining Cosmology with High Convergence Regions\\
in Weak Lensing Surveys}
\author{Sheng Wang\altaffilmark{1,2,3},
Zolt\'{a}n Haiman\altaffilmark{4}, and
Morgan May\altaffilmark{1}}
\altaffiltext{1}{Brookhaven National Laboratory,
Upton, NY 11973--5000, USA} 
\altaffiltext{2}{Department of Physics, Columbia University,
550 West 120th Street, New York, NY 10027, USA}
\altaffiltext{3}{Kavli Institute for Cosmological Physics,
University of Chicago, 933 East 56th Street, Chicago, IL 60637, USA}
\altaffiltext{4}{Department of Astronomy, Columbia University,
550 West 120th Street, New York, NY 10027, USA}
\begin{abstract}
We propose to use a simple observable, the fractional area of ``hot
spots'' in weak gravitational lensing mass maps which are detected with
high significance, to determine background cosmological parameters.
Because these high-convergence regions are directly related to the
physical nonlinear structures of the universe, they derive cosmological
information mainly from the nonlinear regime of density fluctuations.
We show that in combination with future cosmic microwave background
anisotropy measurements, this method can place constraints on
cosmological parameters that are comparable to those from the redshift
distribution of galaxy cluster abundances.  The main advantage of the
statistic proposed in this paper is that projection effects, normally
the main source of uncertainty when determining the presence and the
mass of a galaxy cluster, here serve as a source of information.
\end{abstract}
%
%\keywords{cosmology: theory --- gravitational lensing --- large-scale
%structure of universe --- methods: numerical}
%
\section{Introduction}
\label{sec:introduction}
Weak gravitational lensing (WL), i.e., the coherent distortion of images
of faint distant galaxies by the gravitational tidal field of the
intervening matter distribution \citep{twv90}, has been established as a
powerful cosmological tool [see, e.g., \citet{rev} for a review].  Since
this effect is purely gravitational, it directly probes the matter
distribution along the line of sight, thus providing a way to understand
the nature and the evolutionary history of the universe that is
relatively insensitive to how light or baryons trace dark matter.  Since
the earliest measurements of WL by galaxy clusters \citep{for88,twv90},
WL by large-scale structure, known as ``cosmic shear'', has been
detected based on optical \citep{cs00a,cs00b,cs00c,cs00d} and radio
observations \citep{crh04}.

Future WL surveys, such as the {\it Large Synoptic Survey Telescope}
(LSST)\footnote{www.lsst.org}, will measure the cosmic shear field with
great precision over half the sky.  Recent attention has focused on how
best to extract information from the shear/convergence field in order to
constrain cosmological parameters.  The primary focus so far has been
using the standard two-point statistics \citep{bsbv91,mir91,kai92},
which probes the underlying matter power spectrum in projection.  In
this paper, we propose to use a simple, direct observable: the fraction
of high signal-to-noise ratio ($S/N$) points detected in WL surveys, as
another discriminator of cosmology.  We utilize the results of an N-body
simulation to quantify both the theoretical predictions and the
observational uncertainties of this statistic.

As is well-known, the common two-point statistics do not contain all the
statistical information of the WL convergence field, as the nonlinear
gravitational instability induces non-Gaussian signatures in the mass
distribution and hence in the WL convergence field as well.
Unfortunately, there is no complete statistical analysis in practice for
the underlying matter density field in the nonlinear regime.  Previous
studies of weak lensing statistics on small angular scales,
correspondingly, were restricted to the low-order statistics, using the
halo model [see, e.g., \citet{cs02} for a review], or the ``scaling
ansatz'' \citep{hmkl}, later extended and calibrated by N-body
simulations \citep{jmw95,pd96,smi03}, for the nonlinear evolution of
clustering.  These low-order statistics include: the two-point
correlation function or equivalently the power spectrum, at small scales
\citep{js97}, the three-point correlation function or the bispectrum
\citep{tj03a,tj03b,tj04}, the third-order moment: skewness
\citep{bvm97,js97,hui99} and the fourth-order moment: kurtosis
\citep{tj02}.  An alternative approach is to use the redshift
distribution of nonlinear object abundance \citep{hmh01}, such as
shear-selected galaxy cluster samples \citep{wk03,us,fh07}, using the
Press-Schechter prescription \citep{ps,bcek} with calibration by N-body
simulations \citep{st99,jen01}.  In addition, the ``ratio-statistic''
\citep{jt03,bj04,sk04,hj04,zhs05} has been constructed to use the
geometrical information from tomographic shear/convergence power spectra
on small scales.  These analyses have shown that comparable information
may be contained in the linear and in the nonlinear regime.

The statistics mentioned above are by no means a full characterization
of the convergence field, yet they all face challenges, either from the
theoretical or the observational side, which need to be addressed.
Including higher-order statistics might give a substantial increase in
information, but they are in practice noisy and computationally
intensive.  To take advantage of the synergy between these statistics,
one needs to account for their covariance, which is difficult to
calculate analytically.  One hope is to run large simulations and
directly measure this covariance.  Galaxy cluster samples selected
optically, by their X-ray flux or by their Sunyaev-Zel'dovich effect
signatures, are limited by the uncertain astrophysics when modeling the
mass-observable relations and require ``self-calibration''
\citep{mm03,us,lh05}, as the mass function is exponentially sensitive to
errors of limiting mass.  Shear-selected galaxy clusters, on the other
hand, have the advantage that the selection function can be determined
ab initio by N-body simulations, since only gravity is involved.
However, projection effects result in false detections, missing clusters
\citep{wvm02,hty04,hs05}, and producing significant uncertainty in the
cluster mass derived from the shear signals \citep{mwnl99,dw05},
degrading the cosmological information content.

In this paper, we instead focus on the one-point probability
distribution function (PDF) of the WL convergence field.  There are
three main motivations.  First, the one-point PDF is a simple yet
powerful tool to probe non-Gaussian features, and since the
non-Gaussianity in the convergence field is induced by the growth of
structure, it holds cosmological information
\citep{rkjs99,jsw00,ks00,vmb05}.  These previous works have shown that
the one-point PDF is capable of discriminating cosmologies with
different $\Omega_m$, such as an open cold dark matter (CDM) model, a
flat cosmological constant-dominated ($\Lambda$)-CDM model, and a
standard CDM model.

Second, and the primary motivation of this work, is that the fractional
area statistic we propose in this paper takes into account projection
effects by construction.  Determined by the high-convergence tail of the
PDF, it is an analog of Press-Schechter formalism, thus similar to the
abundance of galaxy clusters but without contamination due to projection
effects.  This statistic also utilizes information mainly from the
nonlinear regime and complements the well-established statistics in the
linear regime. The goal of our work is to give a more quantitative
assessment of the statistical information in the nonlinear regime
(particularly focusing on the properties of dark energy) provided by
this fractional area statistic and its complementarity to other probes
of cosmology, e.g., the cosmic microwave background (CMB) anisotropies.

Third, besides utilizing the cosmic shear field \citep{ztp05} which we
will focus on in this paper, there are several other observational
techniques which could be used to map out the convergence PDF.  For
example, with forthcoming large samples of high redshift supernovae from
LSST-like or Joint Dark Energy Mission (JDEM)-like surveys, one could
measure the magnification distribution of these standard candles due to
lensing by the large scale structure in the foreground, and construct
the convergence PDF \citep{dv06,chh06}; it is also possible to measure
the convergence field through the statistics of cosmic magnification
\citep{j02}, using, for instance, 21cm-emitting galaxies
\citep{zpp05,zpp06}.

The rest of this paper is organized as follows.  Our basic calculational
methodology, including a calibration of the (co)variance using
simulation outputs, is described in \S~\ref{sec:II}.  Results of using
the fractional area statistic for an LSST-like WL survey are presented
in \S~\ref{sec:III}, with discussions of its complementarity to other
dark energy probes, as well as of various uncertainties.  Conclusions
and implications of this work are given in \S~\ref{sec:IV}.  Finally, in
a series of appendices, we show the details of our calculations.
\section{Calculational Method}
\label{sec:II}
\subsection{Convergence Field with Gaussian Smoothing}
Consider some source galaxies detected in a WL survey, with a redshift
distribution of $n(z)$ and a surface density of $n_g=\int n(z)dz$.  The
shear field is measured from the distortion of their images.  A
convergence field, or sometimes called mass map, can be reconstructed
from the shear field \citep{ks93,sk96,bnss96,vbm99}, smoothed over scale
$\theta_G$ with a Gaussian window function
$W_G(\theta)=\exp(-\theta^2/\theta^2_G)/(\pi\theta^2_G)$.  This map is a
sum of the true, smoothed convergence field and the noise field due to
the randomly oriented intrinsic ellipticities of source galaxies:
\be
K(\bt)=\kappa_{(S)}(\bt)+\kappa_{(N)}(\bt).
\label{eq:s+n}
\ee
Both fields are assumed to have zero mean and are statistically
isotropic (ensemble average is same along each line of sight).

Under the assumption that the correlation of intrinsic ellipticities and
the clustering of the source galaxies can be neglected, and no other
systematic errors are present, \citet{van00} has shown that the noise
field $\kappa_{(N)}$ can be modeled as a Gaussian random field with
variance:
\be
\sigma^2_{(N)}\equiv\langle\kappa_{(N)}^2\rangle=
\frac{\sigma^2_\epsilon}{4\pi\theta^2_Gn_g},
\label{eq:varn}
\ee
where $\sigma_\epsilon$ is the root-mean-square value of the intrinsic
ellipticity of source galaxies.  The two-point correlation function of
$\kappa_{(N)}$ induced by the smoothing is
\be
C_{(N)}(\theta)=\sigma^2_{(N)}\exp\left(
-\frac{\theta^2}{2\theta^2_G}\right),
\ee
where $\theta\equiv|\bt_1-\bt_2|$ is the angular separation between two
points in the field.

The true convergence field, on the other hand, is essentially
non-Gaussian.  Its one-point PDF is skew, because $\kappa_{(S)}$ has a
minimum value, corresponding to an totally empty path between the source
and the observer.  The convergence without smoothing, by using Born
approximation, is a weighted projection along a particular line of sight
of the mass density perturbation field \citep{bvm97}:
\be
\kappa(\bt)=\int^{\infty}_0d\chi\,\cw(\chi)\delta(\chi\bt,\chi),
\ee
where $\chi$ is the comoving distance to redshift $z$, and $\delta$ is
the over-density at comoving distance $\chi$.  Taking $\delta=-1$
everywhere along the line of sight, gives the minimum value of
$\kappa_{(S)}$:
\be
\kappa_{(S)\rm min}=\kappa_{\rm min}=-\int^{\infty}_0d\chi\,\cw(\chi),
\label{eq:kmin}
\ee
as the smoothing has no effect on a constant\footnote{Strictly speaking,
this is not true if one measures the minimum value from simulations or
real surveys, due to the finite-sampling effect: it is more rare to have
totally empty lines of sight for the whole smoothing aperture.  This
effect thus results a higher ``$\kappa_{\rm min}$'' for a smoothed
field.}.  For a fixed source plane at redshift $z_s$ corresponding to
$n(z)=n_g\delta(z-z_s)$, the weight function is given by (assuming a
flat universe for simplicity):
\be
\cw(\chi,\chi_s)=\frac{3}{2}\,\Omega_m\left(\frac{H_0}{c}\right)^2
(1+z)\,\chi\left(1-\frac{\chi}{\chi_s}\right)\ch(\chi_s-\chi),
\label{eq:wink}
\ee
where $\chi_s\equiv\chi(z_s)$ denotes the comoving distance to the
source plane and $\ch$ is the Heaviside step function.  Note that this
quantity $\kappa_{\rm min}$, which is a source of the non-Gaussianity of
the convergence field, depends on the source redshift and on cosmology.
The covariance and the correlation function of the smoothed convergence
field, using Limber's approximation, are given in Appendices A and C as
Eq.~(\ref{eq:varkpa}) and (\ref{eq:acfln}).

Points with high $S/N$ in the convergence field are of both experimental
and theoretical interest: not only are they related to the underlying
physical density field at high fidelity, but they capture the important
features of the nonlinear regime as well.  Mathematically, they
correspond to the so-called ``excursion set'', whose properties have
been studied extensively in other similar contexts \citep[most famously
to describe the halo mass function, e.g.,][]{bcek}.  For pedagogical
purposes, let us consider the excursion set $E_\nu$ of the true smoothed
convergence field at redshift $z_s$, defined as the union of all points
with $\kappa_{(S)}>\nu\sigma_{(N)}$ (see \S~\ref{sec:III} for including
the noise field).  The fractional area of $E_\nu$ is then given by
\be
\cf(\nu,z_s)\equiv\frac{A_\nu}{A_{\rm tot}}=
\frac{1}{A_{\rm tot}}\int d^2\bt\,\ch[\kappa_{(S)}(\bt,z_s)-\nu\sigma_{(N)}],
\ee
where $A_\nu$ is the area of the excursion set $E_\nu$, and $A_{\rm
tot}$ is the total sky coverage of a survey.  $\kappa_{(S)}$ is a
statistical quantity, so the fractional area has a mean value of
\be
\langle\cf(\nu,z_s)\rangle=\frac{\langle A_\nu\rangle}{A_{\rm tot}}=
\int_{\nu\sigma_{(N)}}^{\infty}d\kappa_{(S)}\,P_1(\kappa_{(S)}),
\label{eq:mfa}
\ee
where $P_1$ is the one-point PDF of $\kappa_{(S)}$.  The spatial
integration cancels with $A_{\rm tot}$ because of the statistical
isotropy of the field\footnote{Equivalently, we can assume ergodicity,
so that the spatial average is equal to the ensemble average.}.
Similarly, the fractional area has a variance determined by the
two-point PDF, which is given in Appendix C by Eq.~(\ref{eq:tmpcov}).
\subsection{Statistical Properties of the True Convergence Field}
Let us first focus on the one-point PDF of $\kappa_{(S)}$ at different
redshifts: $P_1(\kappa_{(S)},z_s)$, which determines the mean fractional
area $\langle\cf(\nu,z_s)\rangle$.  Several works have used the
``stable-clustering ansatz'' \citep{peebles} to derive the one-point PDF
of the true convergence, with \citep{v00b,mj00,whm02} and without
\citep{v00a} smoothing, calibrated by N-body simulations.  The important
conclusion of these works is that there exists a universal one-point
PDF, well approximated by a two-parameter family: the variance of the
reduced convergence $\langle\kappa^2_{(S)}\rangle/\kappa^2_{\rm min}$
and the minimum convergence $\kappa_{\rm min}$.

In this section, we concentrate on the constraint equations of the
one-point PDF required by the normalization, the mean, and the variance
\citep{do06}.  By normalizing $\kappa_{(S)}$ to
$\kappa'=\kappa_{(S)}/|\kappa_{\rm min}|$, the constraints can be
written as\footnote{These three constraints will be sufficient to
specify a functional form of PDF with no more than three fitting
parameters.  For more complicated models, one needs to go beyond the
variance and specify for the higher-order moments their dependence on
the variance and the minimum.}
\bea
\nn
\int_{-1}^{\infty}d\kappa'\,P_1(\kappa')&=&1,\\
\int_{-1}^{\infty}d\kappa'\,P_1(\kappa')\kappa'&=&0,\\
\nn
\int_{-1}^{\infty}d\kappa'\,P_1(\kappa')\kappa'^2&=&
\langle\kappa_{(S)}^2\rangle/\kappa^2_{\rm min}.
\eea
We consider three different fitting formulae for the one-point PDF of
the convergence field from the literature.  In the next section, we will
use WL simulation outputs to assess the accuracy of the high convergence
tail of these three models.

The first model we consider is the log-normal distribution studied in
\citet{lognor}, and is given by
\bea
&&P_1(\kappa_{(S)})\,d\kappa_{(S)}=\frac{d\kappa_{(S)}}
{|\kappa_{\rm min}|+\kappa_{(S)}}\\
\nn &&\times\frac{1}{\sqrt{2\pi\sigma^2_{LN}}}
\exp\left\{-\frac{[\ln(1+\kappa_{(S)}/|\kappa_{\rm min}|)
+\sigma^2_{LN}/2]^2}{2\sigma^2_{LN}}\right\}.
\eea
Only one parameter,
$\sigma^2_{LN}=\ln{(1+\langle\kappa^2_{(S)}\rangle/\kappa^2_{\rm
min})}$, is needed to satisfy the constraint equations.

The second model is the stretched Gaussian distribution proposed in
\citet{whm02},
\be
P_1(\kappa_{(S)})\,d\kappa_{(S)}=\frac{d\kappa_{(S)}}
{|\kappa_{\rm min}|}C_{\rm norm}\exp\left\{-\left[
\frac{(1+\kappa_{(S)}/|\kappa_{\rm min}|)-\eta_{\rm peak}}
{p(1+\kappa_{(S)}/|\kappa_{\rm min}|)^q}\right]^2\right\}.
\ee
Here we closely follow the original notation, except the PDF is now
written in terms of $\kappa_{(S)}$ instead of $\eta\equiv
1+\kappa_{(S)}/|\kappa_{\rm min}|$.  As stated earlier, the four fitting
parameters $C_{\rm norm}$, $\eta_{\rm peak}$, $p$ and $q$ depend only
upon $\langle\kappa^2_{(S)}\rangle/\kappa^2_{\rm min}$ [see Eq.~(8) of
\citet{whm02}].  The complication of this model is that there is a need
to impose an upper limit of the convergence $\kappa_{\rm max}$ in order
to get the correct variance.

The third model takes the form of a modified log-normal distribution as
proposed in \citet{do06}:
\bea
&&P_1(\kappa_{(S)})\,d\kappa_{(S)}=\frac{d\kappa_{(S)}}
{|\kappa_{\rm min}|+\kappa_{(S)}}\\
\nn &&\times N\exp\left\{
-\frac{[\ln(1+\kappa_{(S)}/|\kappa_{\rm min}|)+\Sigma^2/2]^2
[1+A/(1+\kappa_{(S)}/|\kappa_{\rm min}|)]}{2\Sigma^2}\right\}.
\eea
Given $\langle\kappa^2_{(S)}\rangle/\kappa^2_{\rm min}$, we can uniquely
specify $N$, $\Sigma$ and $A$ by numerically solving the three
constraint equations.  The dependence of these three parameters on the
variance is plotted in Figure~2 of \citet{do06}.

For the analysis in this paper, we emphasize that the cosmological
dependence of these PDFs, i.e., information on dark energy, enters only
through the variance of the reduced convergence\footnote{The universal
halo mass function similarly depends on cosmology only through
$\sigma(M,z)$, the rms value of the matter density fluctuations.
However, one should keep in mind the essential difference that here the
convergence variance is calculated using the nonlinear power spectrum,
whereas for the cluster mass function, the variance is calculated using
the linear matter power spectrum.}:
$\langle\kappa^2_{(S)}\rangle/\kappa^2_{\rm min}$, and through the
minimum: $\kappa_{\rm min}$ as an additional scaling factor.

Once the form of the convergence PDF is specified, the mean fractional
area is determined by its integral [Eq.~(\ref{eq:mfa})].  The
(co)variance of the fractional area for various thresholds and
redshifts, $\mathrm{Cov}[\cf(\mu,z_i),\cf(\nu,z_j)]$, on the other hand,
requires knowing the joint two-point PDF of the convergence field, as we
can see, e.g., in Eq.~(\ref{eq:tmpcov}).  To the best of our knowledge,
among the three cases considered above, the two-point function can be
computed analytically only in the log-normal model; the results are
presented in Appendix C.  Note that we do not use the covariance matrix
of the log-normal model for our actual calculations -- we will instead
utilize simulation outputs to directly measure the (co)variance.  The
results for the log-normal model will be used only for comparison, and
to justify extrapolations from simulations, as we discuss in the next
section.

\begin{figure*}
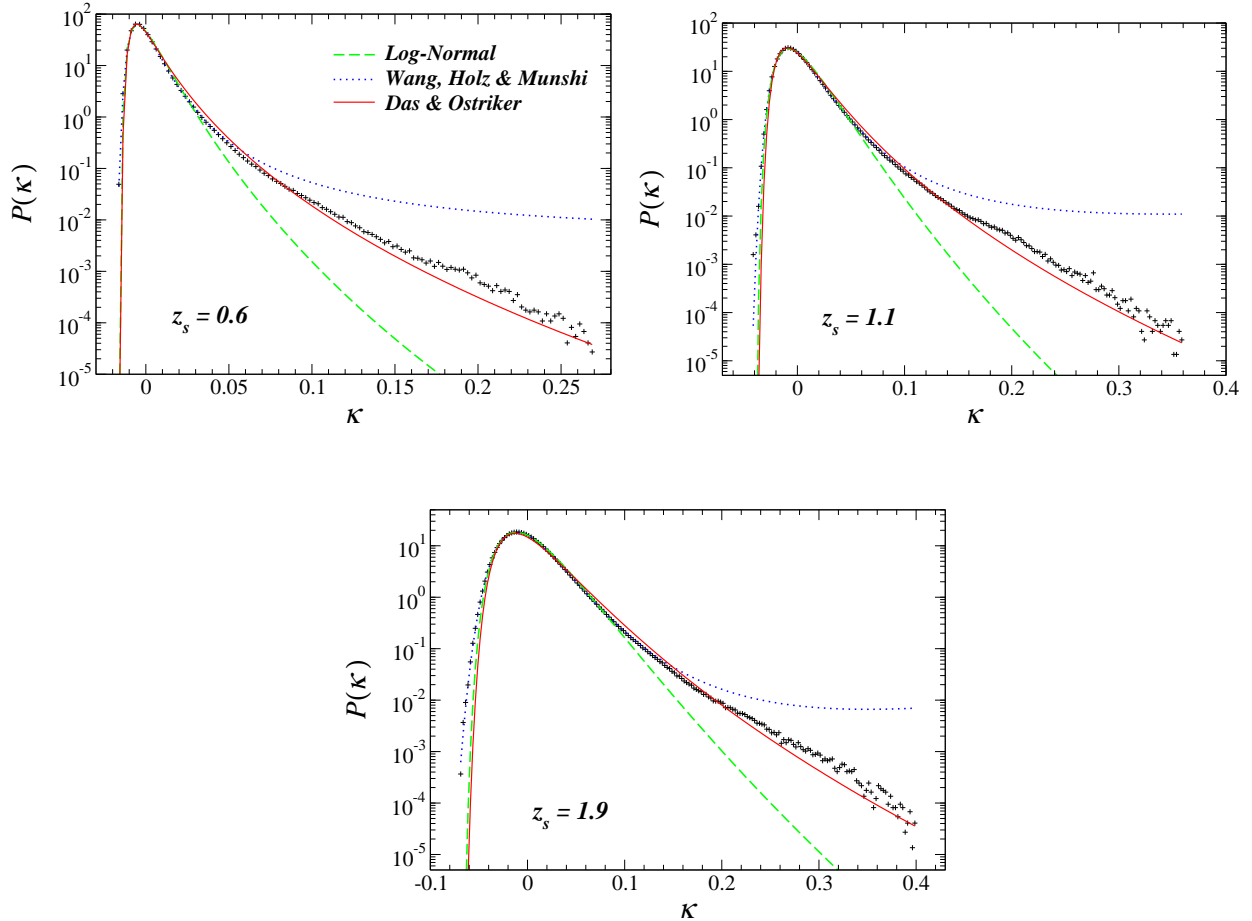
%[htp]
  \begin{tabular}{cc}
    \includegraphics[width=80mm]{f1a}&
    \includegraphics[width=80mm]{f1b}
  \end{tabular}
  \vspace{10mm}\\
  \centering
  \includegraphics[width=80mm]{f1c}
  \caption{\label{fig:1ptpdf}One-point PDFs of the convergence fields.
    Three panels correspond to three distinct redshift bins.  The
    cross symbols are (binned) normalized histograms measured directly
    from the simulation outputs.  The three curves in each panel are
    best-fits to the simulation data using different models as
    described in the text.  Note that the ranges of $x$ and $y$-axes
    are different for each panel.}
\end{figure*}

\subsection{Simulation and Model-Fitting}
The simulation outputs we use are those of \citet{mw05}.  There are a
total of 32 convergence fields from two independent N-body simulation
runs.  Each field is $3\times 3$ deg$^2$, divided into $1024\times 1024$
pixels, so that the angular size of a pixel is $\sim$10.5 arcsec.  There
are three source redshift planes at $z_s=0.6$, 1.1, and 1.9 for each
field, representing the mean redshift of three source galaxy bins of an
LSST-like survey.  The cosmological parameters are adopted from
first-year measurements by the {\it Wilkinson Microwave Anisotropy
Probe} (WMAP)\footnote{map.gsfc.nasa.gov}, as summarized in Table~1 of
\citet{sp03}: a spatially flat $\Lambda$CDM model with a scale-invariant
initial scalar power spectrum ($n_s=1$) and present-day normalization
$\sigma_8=0.9$.  The matter density is $\Omega_m=0.28$ and the baryon
density is $\Omega_b=0.049$.  The Hubble constant is $H_0=70$ km/s/Mpc.

We first apply a Gaussian window function to smooth all the fields, with
$\theta_G$ chosen to be 1 arcmin ($\sim$ 5.7 pixel):
\bea
\nn &&\kappa_{(S)}(i,j)=\displaystyle\sum_{m^2+n^2\leq 32^2}
\kappa(i+m,j+n)\exp\left(-\frac{m^2+n^2}{5.7^2}\right)\\
&&\div\sum_{m^2+n^2\leq 32^2}\exp\left(-\frac{m^2+n^2}{5.7^2}\right).
\eea
$(i,j)$ and $(m,n)$ are integers used to label the coordinates.  For
each pixel, we smooth it with nearby pixels within a circle of a
32-pixel radius which is roughly 6 arcmin, as the Gaussian weighting is
negligible outside.  We discard 32 pixels on each side of the field to
avoid possible edge effects, and use only $960\times 960$ pixel$^2$ in
the center of each smoothed field for the analysis.

In Figure~\ref{fig:1ptpdf}, we show fits to the normalized histograms of
the smoothed convergence fields for the three different PDFs discuss
above (using a bin-size of $\Delta\kappa_{(S)}=0.0025$).  As described
earlier, for all three models, there are only two parameters to fit: the
minimum of the distribution and the variance.  For the log-normal model
and the model proposed by \citet{do06}, all fitting parameters are fixed
once these two quantities are given.  However, for the model proposed by
\citet{whm02}, there is a further complication in determining
$\kappa_{\rm max}$.  For this model, we have therefore relaxed the
assumption above, and allowed all parameters to vary simultaneously,
except we imposed a prior that they lie in the ranges roughly matching
Figure~2 in \citet{whm02}: $0.6<\eta_{\rm peak}<1$, $0<w<0.5$, and
$0.5<q<2.5$.

Recall that the log-normal distribution would be a symmetric parabola on
a log-log plot.  The actual distribution from simulations, however, is
skewed toward the high-$\kappa_{(S)}$ tail [see also \citet{do06}], so
the log-normal PDF under-predicts the tail of the distribution.  In
contrast, the \citet{whm02} model over-predicts the distribution in this
range, as the distribution is cut at $\kappa_{\rm max}$.  Overall, the
model by \citet{do06} goes through the simulation data points remarkably
well at all three redshifts\footnote{One might wonder how one curve in
Figure~\ref{fig:1ptpdf} can lie entirely below another, when the total
area under each of the curves is normalized to unity.  Notice that the
logarithmic y-axis shows a range of many order of magnitudes -- the
tails, where the three models differ, visually dominates these figures,
but the area under these tails is negligible compared to the total area
of the distribution, which lies around the slightly offset peak of each
PDFs.}.

A slightly different expression for $\kappa_{\rm min}$ was given in
\citet{lin07}, taking into account density inhomogeneities by applying
an extended Dyer-Roder formalism \citep{dr72,dr73} to a completely empty
line of sight [also see \citet{ss92}].  The correction to the commonly
used Eq.~(\ref{eq:kmin}) is only large when the sources are at very high
redshifts ($10\%$ for our fiducial cosmology at redshift $z=3$).  In
principle, $\kappa_{\rm min}$ affects the high-tail of the distribution,
since it is used to normalize the universal PDF.  One could also
directly measure this minimum quantity from the simulations.  However,
the low end of the PDF might be really suppressed, because a completely
empty line of sight is extremely unlikely in a finite-size simulation;
the smoothing procedure makes it even less probable.  We have found that
the minimum convergence measured from simulations is nevertheless larger
than theoretical predictions, either using Eq.~(\ref{eq:kmin}) or
\citet{lin07}.  The best-fit $\kappa_{\rm min}$ of the \citet{do06}
model, however, agrees well with the measured minimum value.  This is a
$\sim 20-50\%$ effect, depending on the source redshift [see a more
detailed discussion in \citet{lognor}].  One intriguing question is
which, among the minimum values mentioned above, could lead to a more
universal PDF? Particular, for the purposes of this paper, which one
leads to the most universal high-tail distribution?  We hope to
investigate this question in future work, using larger-volume
simulations with various cosmologies.  In this paper, we simply use
Eq.~(\ref{eq:kmin}) to calculate the derivatives of the fractional areas
with respect to the cosmological parameters.

To conclude, for the purpose of taking derivatives in the Fisher matrix
analysis (see below), we adopt the universal one-point PDF provided by
\citet{do06}, with the variance of the convergence field computed by
using the nonlinear matter power spectrum from \citet{smi03} as given by
Eq.~(\ref{eq:varkpa}) and the minimum as given by Eq.~(\ref{eq:kmin}).

Next, we measure directly from the simulation outputs the covariance
matrix of the fractional areas with various thresholds and redshifts:
${\rm Cov}[\cf(\mu,z_i),\cf(\nu,z_j)]$.  The variance is given by
\bea
\nn &&\mathrm{Var}[\cf(\nu,z_s)]=\frac{1}{N_f-1}\\
&&\times\left[\displaystyle
\sum_{m=1}^{N_f}\left(\frac{N^{(m)}(\nu,z_s)}{N_p}\right)^2-\frac{1}{N_f}
\left(\sum_{m=1}^{N_f}\frac{N^{(m)}(\nu,z_s)}{N_p}\right)^2\right].
\eea
Here $N^{(m)}(\nu,z_i)$ is the total number of pixels with signal above
the threshold $\nu\sigma_{(N)}$, in the $m$-th field at redshift $z_s$;
$N_f=32$ is the total number of fields and $N_p=960^2$ is the total
number of pixels in each field.  Similarly, the covariance is given by
\bea
&&\mathrm{Cov}[\cf(\mu,z_i),\cf(\nu,z_j)]=\frac{1}{N_f-1}\\
\nn &&\times\left[\displaystyle\sum_{m=1}^{N_f}
\frac{N^{(m)}(\mu,z_i)}{N_p}\frac{N^{(m)}(\nu,z_j)}{N_p}
-\frac{1}{N_f}\sum_{m=1}^{N_f}\frac{N^{(m)}(\mu,z_i)}{N_p}
\sum_{m=1}^{N_f}\frac{N^{(m)}(\nu,z_j)}{N_p}\right].
\label{eq:simcov}
\eea

To extrapolate the results we obtain from a 9 deg$^2$ simulation field
to a 20,000 deg$^2$ LSST-like survey, we also measure the covariance
matrix as a function of the field size.  To do this, we divide each
field into several smaller sub-fields.  The following cases are
considered: $N_p=480^2$, $240^2$, $120^2$, $60^2$, and $30^2$.  For each
case, we fix $N_f=32$.  For example, when $N_p=480^2$, we first use the
upper left $480^2$ pixel sub-field of all 32 fields to measure the
covariance using Eq.~(\ref{eq:simcov}).  We repeat the same procedure
for the other three sub-fields, and then take the average of the four
results.

\begin{figure*}%[htp]
  \begin{tabular}{cc}
    \includegraphics[width=80mm]{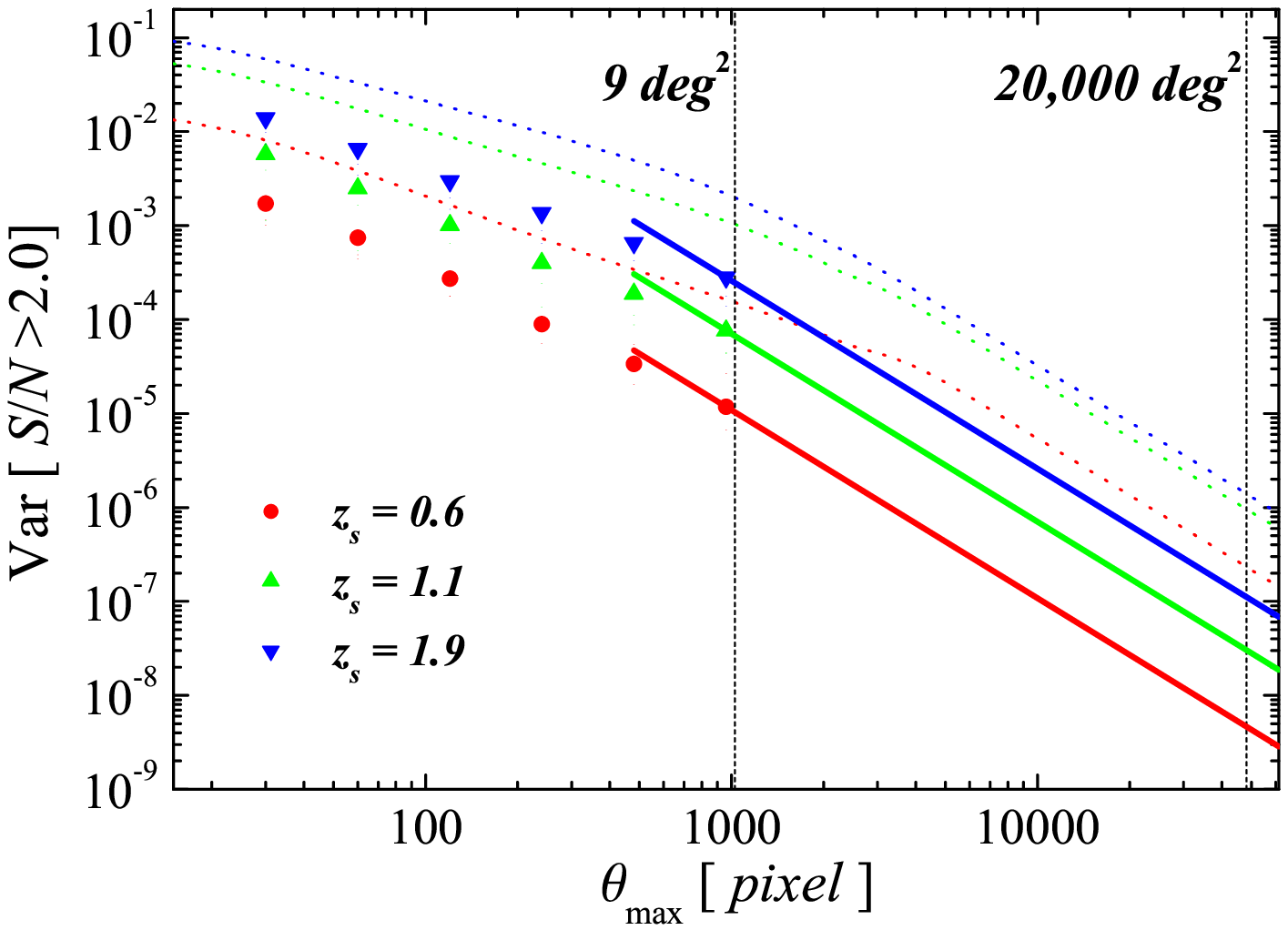}&
    \includegraphics[width=80mm]{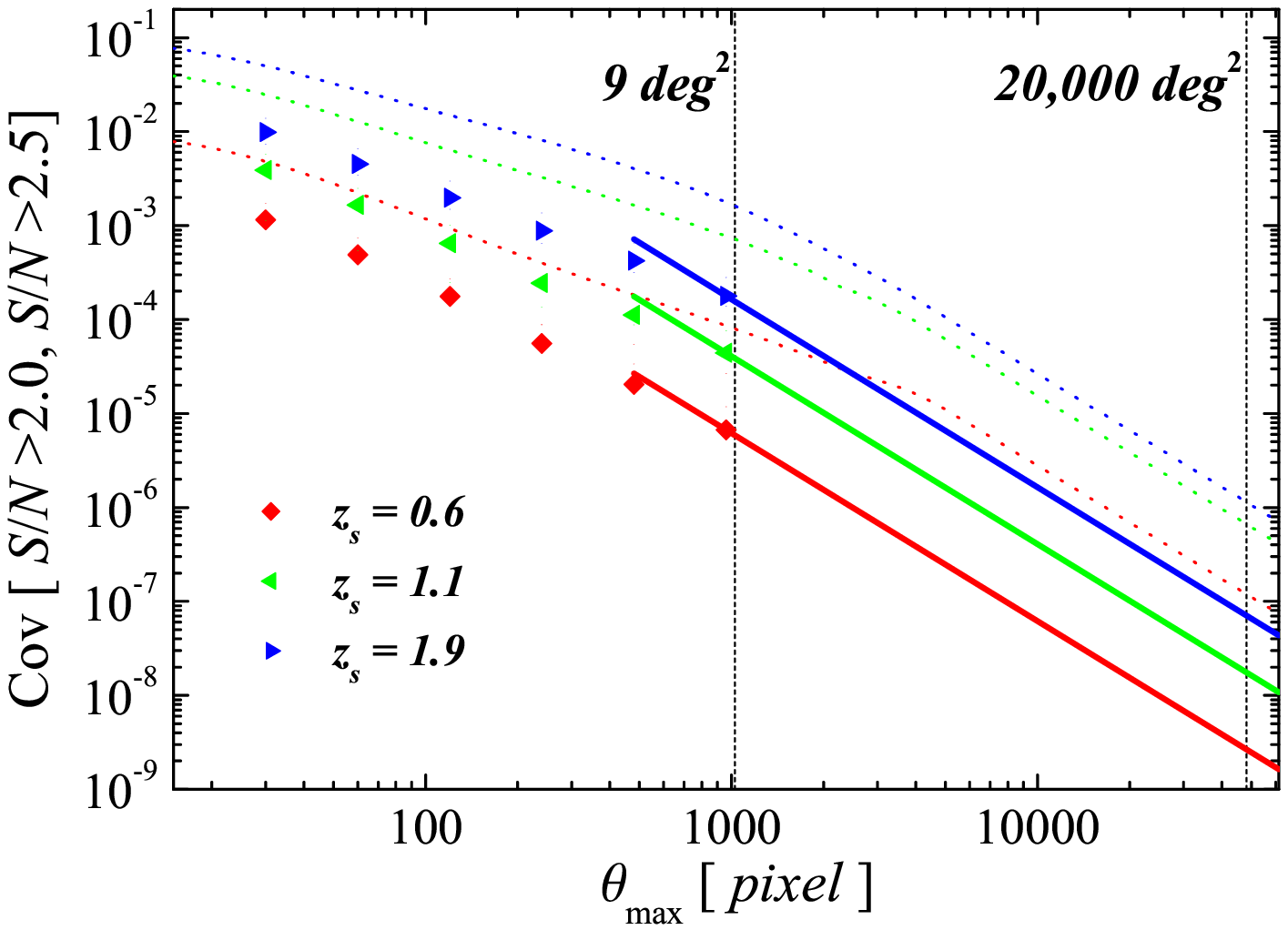}
  \end{tabular}
  \vspace{10mm}\\
  \centering
  \includegraphics[width=80mm]{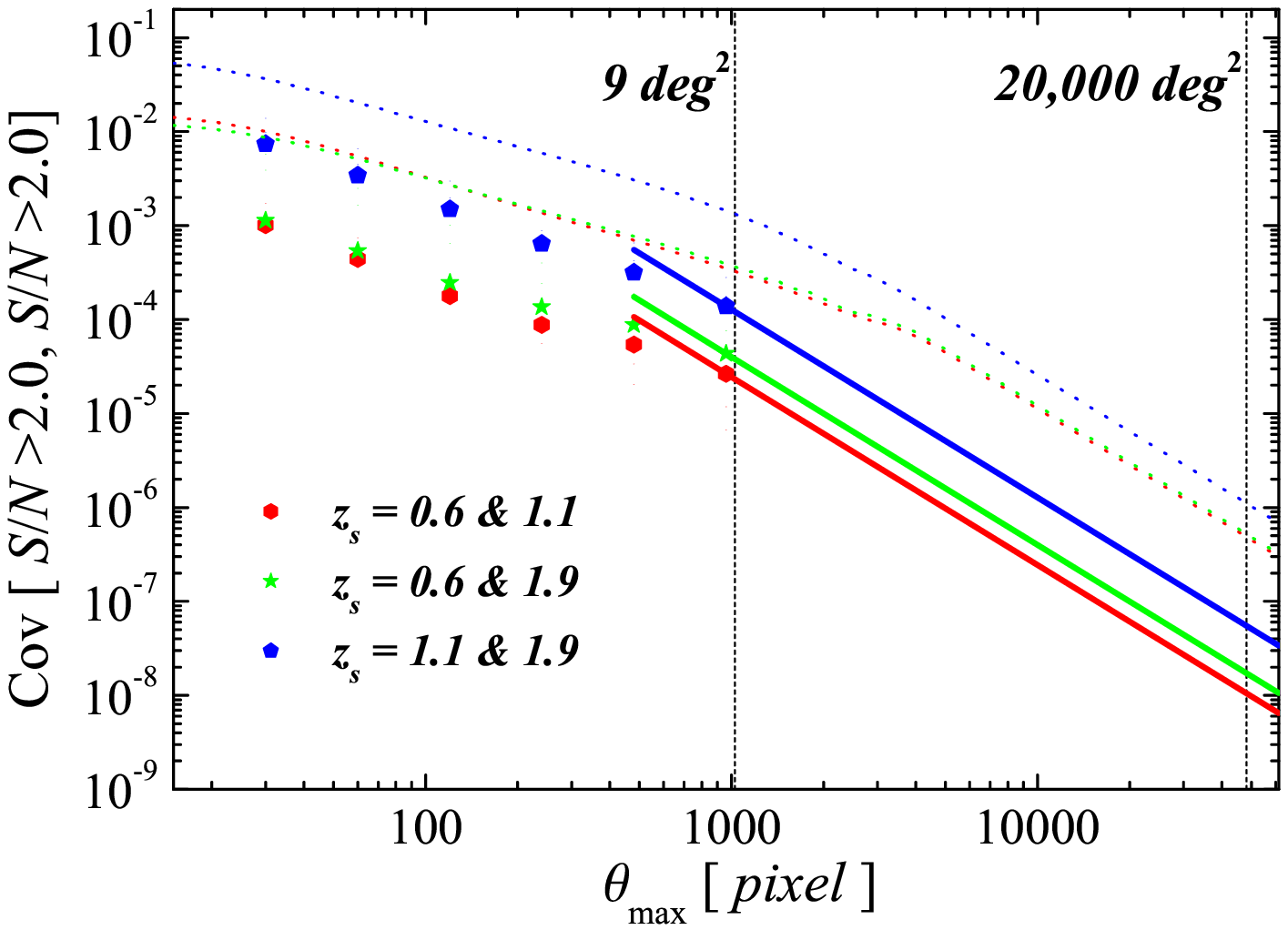}
  \caption{\label{fig:cov}The top left panel shows the variance
    $\mathrm{Var}[\cf(\mu=2,z_i)]$ within three distinct redshift bins;
    the top right panel shows the covariance
    $\mathrm{Cov}[\cf(\mu=2,z_i),\cf(\nu=2.5,z_i)]$ within the same
    three redshift bins.  The bottom panel shows
    $\mathrm{Cov}[\cf(\mu=2,z_i),\cf(\nu=2,z_j)]$ across three different
    pairs of redshift bins.  All symbols in each panel show results
    measured directly from the simulation outputs.  The solid curves
    show the estimate of the covariance matrix through
    Poisson-extrapolation, which we use in the actual calculations.  The
    dotted curves are calculated using the log-normal model, which are
    plotted only for comparison.}
\end{figure*}

In Figure~\ref{fig:cov}, we plot the (co)variance of the fractional
area, as a function of scale, for different thresholds and redshift bins
measured from the simulation outputs.  For small fields, we effectively
have more realizations to measure the covariance matrix, for which we
plot the average.  For comparison, we also plot, as dotted curves, the
values calculated using the log-normal model.

One expects that when the size of the field $\theta_{\rm max}$ becomes
larger than the correlation length of these ``biased regions'' in the
convergence field, the (co)variance scales with respect to the survey
size at least as steeply as $\propto A^{-1}_{\rm tot}$ (coinciding with
a Poisson-like scaling; see Appendix C).  On smaller scales, the slope
is generally shallower due to correlations.  For example, we can observe
this trend analytically in the log-normal case -- the covariance is
given by a Poisson expression [Eq.~(\ref{eq:cov})] for large scales and
an additional non-Poisson correction term [Eq.~(\ref{eq:covd})].  The
simulation points in Figure~\ref{fig:cov} show that for low-redshift
bins, the scale of 1024 pixel (3 deg) is already within the Poisson
region.  For higher redshift bins, the slope at 1024 pixel is shallower
than Poisson.  However, the slope does agree well with the log-normal
model, which predicts that the scale of 1024 pixel is very close to the
edge of the Poisson region.  Thus, for our Fisher matrix analysis, we
estimate the covariance matrix of the fractional area for an LSST-like
survey by Poisson-scaling the measured value from the simulation outputs
with the largest field size of $N_p=960^2$ pixel.  This extrapolation to
20,000 deg$^2$ is shown explicitly in Figure~\ref{fig:cov} as the solid
curves.  The justification for this extrapolation, using the log-normal
model, is admittedly still somewhat heuristic, but measurements of the
covariance out to larger scales require larger size simulations; we
defer this to future work.

\subsection{Random Ellipticity Noise}
In order to incorporate the presence of noise into our forecasts, we
start with the simplest case: modeling the noise due to the intrinsic
ellipticity as a two-dimensional Gaussian random field that is
uncorrelated with the signal \citep{van00,jv00}.  We use a Gaussian
random number generator with zero-mean and a variance of $\sigma_{(N)}$
to generate noise fields and co-add onto the original convergence
fields.  We adopt $\sigma_\epsilon=0.3$, $\theta_G=1$ arcmin, and
$n_g=40/3$ arcmin$^{-2}$ (the denominator ``3'' reflects the redshift
binning) in Eq.~(\ref{eq:varn}) for an LSST-like survey, thus
$\sigma_{(N)}\sim 0.023$.  We then repeat the same Gaussian smoothing
procedure and measure the covariance matrix of the fractional area.  In
this case, the one-point PDF is simply a convolution of the original
distribution and the Gaussian PDF.

Modeling of the noise field as above might be overly simplified, even in
the case when all other systematics are absent.  This is because
\citet{van00} only worked in the WL limit, in which the estimator of the
convergence is safely linear.  We here typically use a signal threshold
of several times $\sigma_{(N)}$, reaching values of order $\sim 5-10\%$.
Construction of the convergence field from the observed reduced shear is
nonlinear, and therefore gets non-negligible higher-order corrections,
which will complicate the expression for the convergence error.  Besides
the intrinsic ellipticity, there will be other systematic effects, due
to the point spread function, mirror distortion, etc, which will
increase the scatter, and may produce an unknown bias in measuring the
cosmic shear.

In our analysis, we adopt an approach similar to what has been done in
the literature for cluster counts in order to account for the
uncertainties of the mass-observable relations \citep{mm03,us,lh05}.  We
still make the simple assumption that the overall convergence noise is
Gaussian and uncorrelated with the signal, but with unknown variance and
bias, which will be included as additional nuisance parameters that are
fit by the survey itself, simultaneously with the cosmology parameters
(``self-calibration'').  We assign the variance and bias in each
redshift bin as free parameters, and their fiducial values are
$\{\sigma_{(N)},\kappa_{\rm bias}\}=\{0.023,0\}$, chosen to be same for
all redshift bins.  For WL power spectrum and bispectrum tomography,
\citet{htbj06} have shown that there is sufficient information in these
statistics themselves to successfully self-calibrate additive and
multiplicative shear systematics.  As we will see below, the nuisance
parameters we consider here likewise do not significantly degrade the
cosmological constraints derived from the one-point statistic, provided
we have some external knowledge of what the intrinsic ellipticity
distribution is.  The reason for this is essentially that the PDF
contains both the tomographic and the shape information (similarly to
the case of cluster counts, where the nuisance parameters cannot
simultaneously mimic changes in both the shape and the $z$-dependence of
the mass function caused by variations in cosmology).
\subsection{Error Forecasts}
The usual Fisher matrix technique \citep{tth97} is employed to assess
the cosmological sensitivity of the fractional area statistic.  This
method allows a quick exploration of the parameter space and gives a
lower bound to the statistical uncertainty of each model parameter to be
fit by future experimental data.

We consider a WL survey with specifications similar to that planned for
LSST: a sky coverage of $A_{\rm tot}=20,000$ deg$^2$, source galaxies
with the redshift distribution adopted from Eq.~(18) in \citet{sk04},
normalized to a total surface number density of $n_g=40$ arcmin$^{-2}$,
and the intrinsic ellipticity dispersion $\sigma_\epsilon=0.3$.  The
galaxies are divided into three bins with mean redshift $z_s=0.6$, 1.1
and 1.9 so that each bin contains the same number of galaxies.  The
Gaussian smoothing is taken over the scale $\theta_G=1$ arcmin within
the nonlinear regime.  Seven different $S/N$ thresholds, from $\nu=2.0$
to $5.0$ in increment of $\Delta\nu=0.5$, are considered simultaneously
for each redshift bin to utilize the information contained in the shape
of the PDF (see \S~\ref{sec:III} below for discussions of different
choices of smoothing scale and $S/N$ thresholds).

The full Fisher matrix is given by Eq.~(15) of \citet{tth97}:
\be
F_{\alpha\beta}=\frac{1}{2}\mathrm{Tr}\left[
{\bf C}^{-1}{\bf C}_{,\alpha}{\bf C}^{-1}{\bf C}_{,\beta}+
{\bf C}^{-1}\left({\bcf}_{,\alpha}{\bcf}^t_{,\beta}+
{\bcf}_{,\beta}{\bcf}^t_{,\alpha}\right)\right]~,
\ee
where ${\bcf}$ is the list of mean fractional areas above different
thresholds and at different redshifts, arranged into a vector, and ${\bf
C}$ is the covariance matrix between them.  We have used the standard
comma notation for derivatives.

In this paper, we neglect the cosmological dependence of the covariance
matrix, i.e. the sample variance\footnote{Just like the full Fisher
matrix, the sample variance matrix is positive definite, because it is a
special case when the observables have mean values of, for example,
zero.  Thus we expect better constraints if this term was included,
since it contains information from two-point statistics of these WL
``biased'' regions.  This is again analogous to the case of galaxy
clusters, where \cite{lh05} have shown that sample-variance helps with
self-calibration.  We hope to explore this in a future work.}.  It is
straightforward to spell out the last two terms, which gives
\bea
\nn
F_{\alpha\beta}&=&\frac{1}{2}\mathrm{Tr}\left[{\bf C}^{-1}
\left({\bcf}_{,\alpha}{\bcf}^t_{,\beta}+
{\bcf}_{,\beta}{\bcf}^t_{,\alpha}\right)\right]=
{\bcf}^t_{,\alpha}{\bf C}^{-1}{\bcf}_{,\beta}\\&=&
\sum_{i,j}\sum_{\mu,\nu}\frac{\partial\langle\cf(\mu,z_i)\rangle}{\partial p_\alpha}
({\bf C}^{-1})_{(\mu,i)(\nu,j)}\frac{\partial\langle\cf(\nu,z_j)\rangle}{\partial p_\beta},
\label{eq:fisher}
\eea
where $p_\alpha$ includes the cosmological parameters of interest, as
well as possible nuisance parameters.  The set of different redshift
bins and $S/N$ thresholds are labeled by $i$ or $j$, and $\mu$ or $\nu$,
respectively.

The covariance matrix ${\bf C}$ is estimated by extrapolation from
simulations, as discussed above. Note that in Eq.~(\ref{eq:fisher}),
this covariance matrix is needed only for the fiducial cosmology.  The
subscript ``$(\mu,i)(\nu,j)$'' denotes the corresponding matrix element.
The choice of cumulative $S/N$ thresholds instead of independent bins is
made to simplify the calculations in the Appendices.  Note that the
off-diagonal terms of the covariance matrix, reflecting the
cross-correlations between different redshift bins and $S/N$ thresholds,
are nonzero so that double-counting information from overlapping lens
redshift and $S/N$ ranges is avoided.

We consider a spatially flat cosmological model with seven parameters.
The fiducial values are close to those adopted in the simulations:
$\{\Omega_{\rm DE},w_0,w_a,\Omega_mh^2,\Omega_bh^2,\sigma_8,n_s\}=$
$\{0.72,-1,0,0.137,0.024,0.9,1\}$ (note a higher $\sigma_8$ than WMAP
three-year result).  The dark energy equation of state is parametrized
as \citep{cp01,lin03}
\be
w(a)=w_0+w_a(1-a)=w_0+w_a\frac{z}{1+z}.
\label{eq:wz}
\ee

CMB anisotropies have played a crucial role in firmly establishing the
current ``standard model'' of cosmology \citep{sp03,sp07}.  To take
advantage of the future high precision data which will be available from
planned CMB anisotropy measurements, a survey with specifications
similar to the {\it Planck surveyor}
(Planck)\footnote{www.rssd.esa.int/index.php?project=PLANCK} is
considered.  The Fisher matrix for the CMB temperature and polarization
anisotropies is constructed as given in \citet{us}.

\begin{table*}[htb]\small
\caption{Calibrated Cosmological Parameter Constraints from LSST using
  the fractional area statistic and adding Planck priors:
  $\Delta\Omega_m h^2=0.0012$, $\Delta\Omega_b h^2=0.00014$, and
  $\Delta n_s=0.035$.  For the pessimistic scenario, we adopt $1\%$
  priors on the additive and multiplicative errors already achieved by
  current surveys.  For the optimistic scenario, we adopt $0.01\%$ and
  $0.05\%$ priors on the additive and multiplicative error respectively,
  which is the goal of future surveys.\label{tab:all}}
\begin{center}
\hbox to \hsize{\hfil\begin{tabular}{|c|c|c|}
\hline
Parameter Constraints&LSST ($\cf$)&LSST ($\cf$) + Planck (priors)\\
\hline
\hspace{5pt}{\bf Pessimistic}       &        &        \\
\hspace{5pt}$\Delta w_p$            & 0.094  & 0.022  \\
\hspace{5pt}$z_p$                   & 0.50   & 0.94   \\
\hspace{5pt}$\Delta w_0$            & 0.55   & 0.12   \\
\hspace{5pt}$\Delta w_a$            & 1.6    & 0.25   \\
\hspace{5pt}$\Delta\Omega_{\rm DE}$ & 0.046  & 0.0095 \\
\hspace{5pt}$\Delta\sigma_8$        & 0.047  & 0.0080 \\
\hline
\hspace{5pt}{\bf Optimistic}        &        &        \\
\hspace{5pt}$\Delta w_p$            & 0.028  & 0.012  \\
\hspace{5pt}$z_p$                   & 0.60   & 0.60   \\
\hspace{5pt}$\Delta w_0$            & 0.16   & 0.043  \\
\hspace{5pt}$\Delta w_a$            & 0.42   & 0.11   \\
\hspace{5pt}$\Delta\Omega_{\rm DE}$ & 0.015  & 0.0038 \\
\hspace{5pt}$\Delta\sigma_8$        & 0.031  & 0.0029 \\
\hline
\end{tabular}\hfil}
\end{center}
\end{table*}
\begin{figure*}
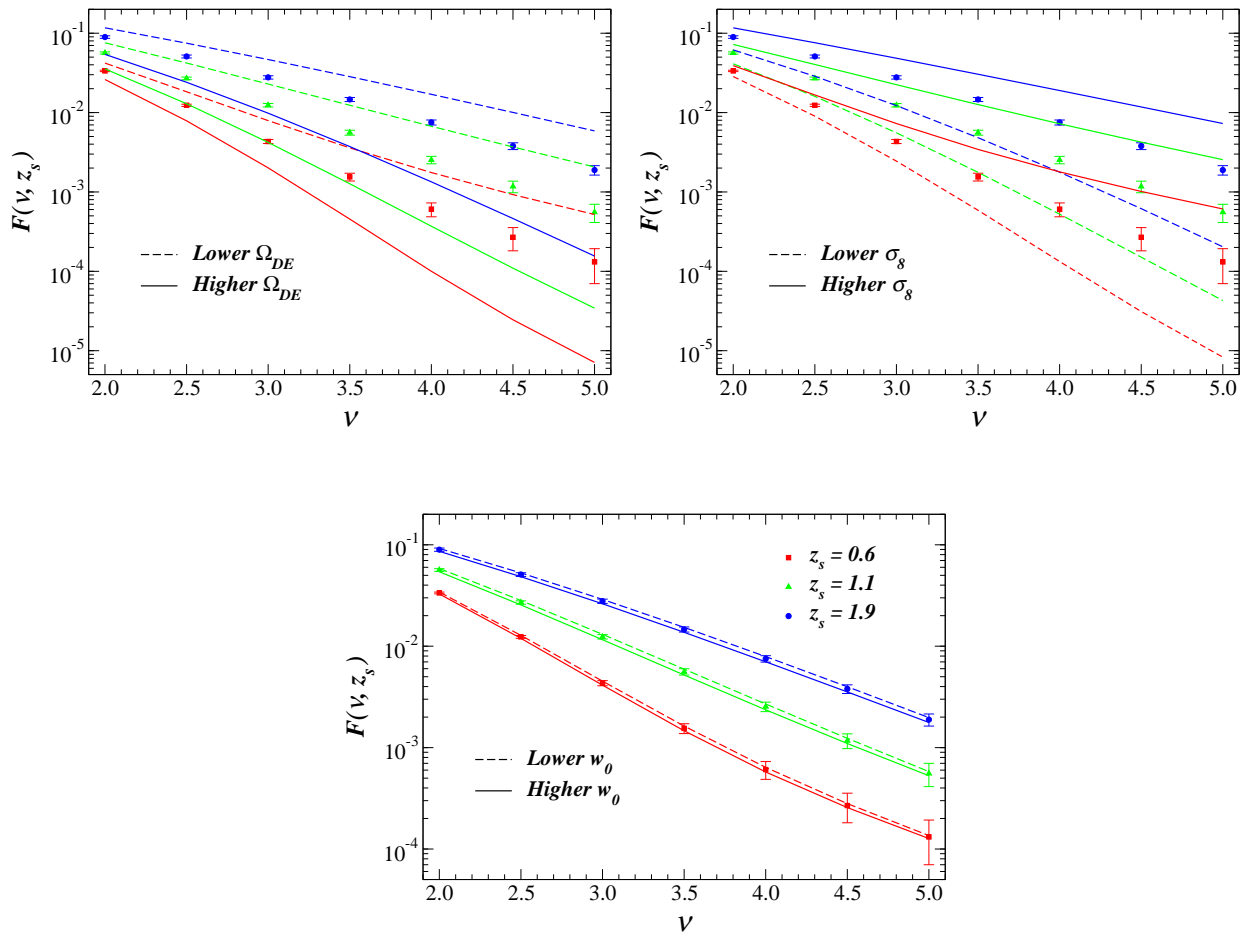
%[htp]
  \begin{tabular}{cc}
    \includegraphics[width=80mm]{f3a}&
    \includegraphics[width=80mm]{f3b}
  \end{tabular}
  \vspace{10mm}\\
  \centering
  \includegraphics[width=80mm]{f3c}
  \caption{\label{fig:dif}Mean fractional area $\langle\cf\rangle$ for
    different S/N threshold $\nu$, different redshift $z_s$, and its
    dependence on various cosmological parameters.  The symbols in
    each panel show $\langle\cf\rangle$ in the fiducial cosmology
    (with the intrinsic ellipticity noise taken into account), with
    the error bars measured directly from the simulations.  Note that
    the error bars are correlated, and they are also enlarged by a
    factor of 10 to be clearly visible.  The solid (dashed) curves
    show results when the given cosmological parameter is $20\%$
    higher (lower) than its fiducial value.}
\end{figure*}
\begin{figure*}
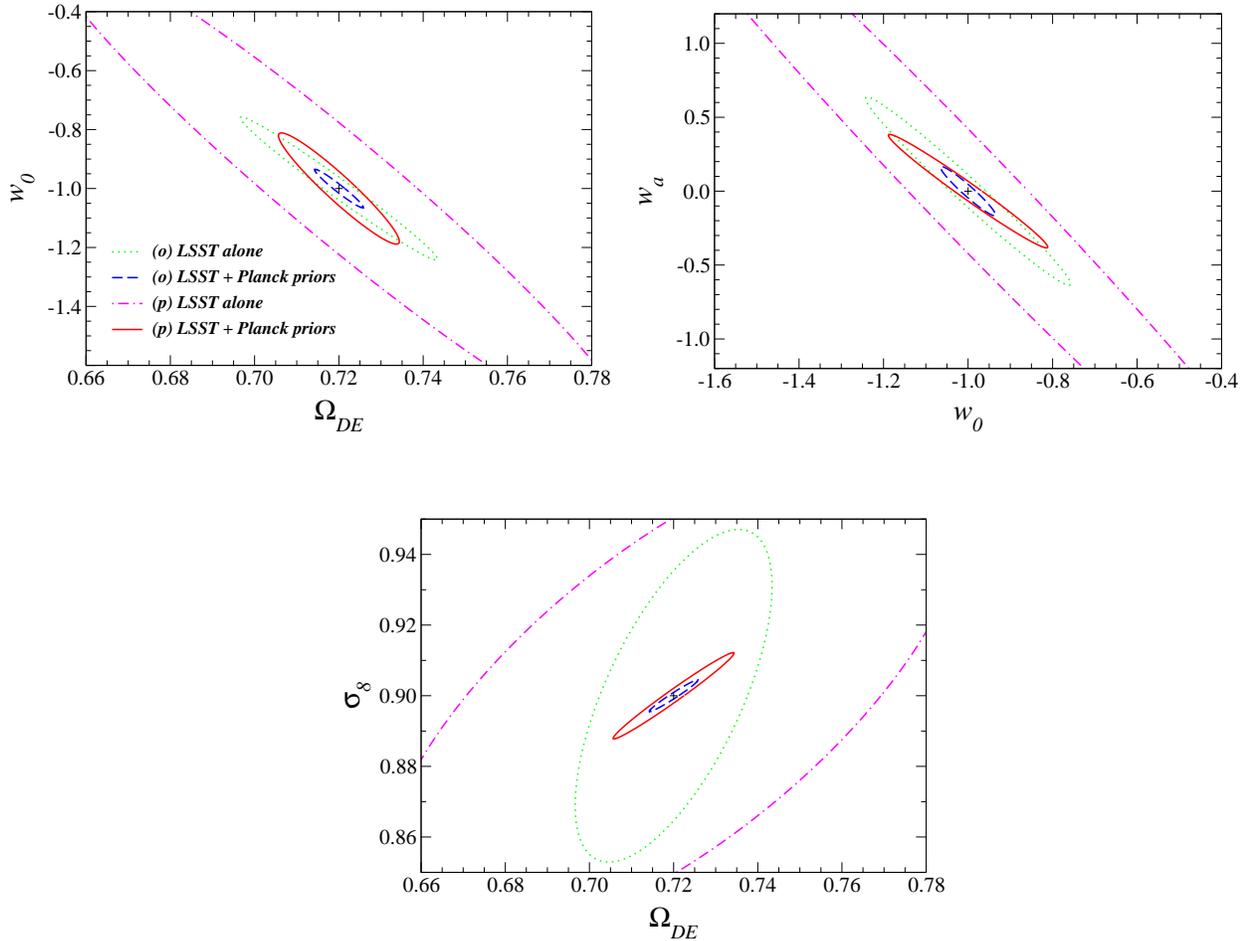
%[htp]
  \vspace{5mm}
  \begin{tabular}{cc}
    \includegraphics[width=80mm]{f4a}&
    \includegraphics[width=80mm]{f4b}
  \end{tabular}
  \vspace{10mm}\\
  \centering
  \includegraphics[width=80mm]{f4c}
  \caption{\label{fig:ell}Constraints on various cosmological parameters
    from an LSST-like weak lensing survey.  The $68\%$ C.L. contours in
    all panels show reasonable optimistic and pessimistic estimates
    including systematic errors for an LSST-like survey, with or without
    priors from a Planck-like CMB survey.  The pessimistic scenario
    corresponds to the additive and multiplicative errors already
    achieved by current surveys.  The optimistic scenario corresponds to
    the accuracy goal of future surveys, such as LSST.  Different contours
    are for LSST alone (dotted curve) and LSST with priors from Planck
    (dashed curve) in the optimistic scenario; LSST alone (dot-dashed
    curve) and with Planck priors (solid curve) in the pessimistic
    scenario.  The contours are marginalized over all other relevant
    parameters.}
\end{figure*}

\section{Results and Discussion}
\label{sec:III}
Before we present the results, it is useful to keep several numbers in
mind.  The variance of the noise field $\sigma_{(N)}\sim 0.023$ due to
the intrinsic ellipticities of the source galaxies is the same for each
redshift bin; the variance of the true smoothed convergence field
$\sigma_{(S)}$ is $\sim 0.009$, 0.017 and 0.025 for $z_s=0.6$, 1.1 and
1.9 respectively.  The lowest threshold $2\sigma_{(N)}$ is chosen so
that the points considered are all in the tail of the underlying PDF:
$\kappa_{(S)}(\bt)>\sigma_{(S)}$.  The reason for restricting ourselves
to the tail instead of using the full PDF is that the points which lie
within $1\sigma_{(S)}$ of the mean probe the linear regime, where the
shear field can be well approximated as Gaussian. In this regime the
power spectrum, which has already been studied extensively, should
contain essentially all of the statistical information.

Figure~\ref{fig:dif} shows how the fractional area statistic depends on
the cosmological parameters $\Omega_{\rm DE},\sigma_8$ and $w_0$.  The
data points in each panel are the fractional areas above different
thresholds for each redshift bin in the fiducial cosmology.  The lines
are calculated by varying the parameter in question by $\pm 20\%$ (and
leaving all other parameters unchanged).  For reference, we have also
shown, as the error bars, the variance measured directly from the
simulations.  The dependencies revealed on this figure make intuitive
sense: decreasing $\Omega_{\rm DE}$ (or equivalently, increasing
$\Omega_{m}$), increasing $\sigma_8$, or making $w_0$ more negative all
affect the growth of structures in the direction of having larger
amplitudes, which increase the shear.

Table~\ref{tab:all} summarizes our main results for an LSST-like survey
alone, as well as its combination with Planck.  We follow the
``self-calibration'' approach, in which we assume $\sigma_{(N)}$ and
$\kappa_{\rm bias}$ are unknown, and fit separately in each redshift
bin.  To account for the uncertainties of the six extra nuisance
parameters, instead of three $S/N$ bins, we utilize seven bins extending
from $\nu=2.0$ to $\nu=5.0$ in increments of $\Delta\nu=0.5$.  We
consider two scenarios for the convergence systematics: the top half of
Table~\ref{tab:all} shows a rather pessimistic projection of the survey,
where we adopt $1\%$ priors\footnote{We give equal prior to each of the
redshift bin.} on $\sigma_{(N)}$ and $\kappa_{\rm bias}$, corresponding
to the additive and multiplicative errors already achieved by current
surveys \citep{fu08}; the bottom half of Table~\ref{tab:all} represents
a more optimistic projection of an LSST-like survey, where we adopt a
$0.01\%$ prior on $\sigma_{(N)}$ and a $0.05\%$ prior on $\kappa_{\rm
bias}$, which is the accuracy goal of future surveys.  Note that in the
last column, we adopt a conservative approach and only include
independent priors on $\Omega_m h^2$, $\Omega_b h^2$, and $n_s$ (the
diagonal approximation instead of the full CMB Fisher matrix).  These
three parameters are expected to be constrained by Planck alone to
$0.9\%$, $0.6\%$ and $0.4\%$ respectively \citep{hu02}.

As noted above, the cosmological dependence of the fractional areas,
similar to galaxy cluster abundance, comes in through two quantities:
(1) the variance of the reduced convergence,
$\langle\kappa^2\rangle/\kappa^2_{\rm min}$, thus most sensitive to the
amplitude of the density fluctuation $\sigma_8$; (2) the minimum value
of the convergence, $\kappa_{\rm min}$, thus most sensitive to the
matter content $\Omega_m$, or $\Omega_{\rm DE}$ as a flat universe is
assumed.  However, the fractional-area statistic also provides good
constraints on the dark energy equation of state, for the following
reason. With Planck priors for $\Omega_m h^2$, $\Omega_b h^2$, and
$n_s$, which are all below the $1\%$ level, an LSST-like survey can
determine $\Omega_{\rm DE}$ and $\sigma_8$ very well (to a few percent
accuracy) by using the fractional area statistic, which in turn breaks
the degeneracies in the growth of structure between these two parameters
and the dark energy equation of state parameters: $w_0$ and $w_a$.

We follow the {\it Dark Energy Task Force} (DETF) report \citep{detf06}
and calculate the pivot scale-factor $a_p$, or equivalently the pivot
redshift $z_p$, where the dark energy equation of state $w(a)$ is best
constrained.  In the pessimistic case, we find $\Delta w_p=0.022$ with
$a_p=0.52$ ($z_p=0.94$) for LSST with Planck prior, where $w_p\equiv
w(a_p)$.  This is improved to $\Delta w_p=0.012$ with $a_p=0.63$
($z_p=0.60$) in the optimistic case.  The corresponding ``figure of
merit''\footnote{Note that the DETF has included and marginalized over
the curvature of the universe in their analysis, whereas we have assumed
a flat universe.  But we expect minor degradation of this ``figure of
merit'' once we also include the Planck prior on the curvature.} is
$(\Delta w_p \Delta w_a)^{-1}\sim 180$ for the pessimistic projection,
which is at the interesting level [compared to the various stage-III
experiments; see \citet[][page 77]{detf06}], and 760 for the optimistic
scenario; the best-constrained pivot point is also at somewhat higher
redshift than other proposed probes, which gives a different degeneracy
direction on the $w_0$-$w_a$ plane, so we expect good synergy between
the fractional area statistic and other dark energy probes.  It is also
different from the main degeneracy direction for CMB anisotropies, the
angular diameter distance degeneracy, which gives $a_p=0.72$
($z_p=0.38$) for our fiducial cosmology.

In Figure~\ref{fig:ell}, we show the two-dimensional marginalized error
contours of various cosmological parameters.  As shown in the bottom
panel, there is still a strong degeneracy between $\Omega_{\rm DE}$ (or
equivalently $\Omega_m$, as we have assumed a flat universe) and
$\sigma_8$ in the self-calibration case.  Consequently, this degrades
the constraints of $w_0$ and $w_a$ by large factors, as shown in the
upper panels of Figure~\ref{fig:ell}.  We have found that, comparing the
pessimistic scenario with the optimistic one, there is respectively, a
factor of 2 and 2.5 degradation for the constraint on $w_p$ and $w_a$
(when Planck priors are added; the degradation is more significant for
the convergence-statistic only case).  As illustrated by \citet{htbj06},
the prior information of the intrinsic ellipticity noise can help to
restore the dark energy constraints.  We also note that the prospects
for self-calibration of systematic errors of an LSST-like WL survey
would be further improved if the extra information from the tomographic
power spectrum or bispectrum were added in the analysis.

The numbers quoted above should be interpreted as the lower bound of the
cosmological sensitivities one can get from an LSST-like survey using
the statistics described (with Planck priors).  In reality, various
systematics have to be considered, which we discuss next.

For an LSST-like survey, most of the source galaxies will only have
photometric redshift information.  To check the sensitivity of the
fractional area statistic to the redshift uncertainties, we calculated
the change in the source redshift which would cause a shift of the
fractional area by more than its $1\sigma$ error.  It is estimated using
$\Delta z_s=(d\langle\cf\rangle/dz_s)^{-1}\times$
$\sqrt{\mathrm{Var}[\cf(z_s)]}$, where $d\langle\cf\rangle/dz_s$ is the
derivative of the mean fractional area with respect to source redshift.
We have found that the systematic error (bias) of photometric redshift
of the source galaxies needs to be below $0.002$ for low redshift and
$0.01$ for $z>2$, which is within the expectations of LSST [see
assumption in \citet{mhh06,htbj06}].  The requirement on the photometric
redshift scatter will not be a major problem, because many source
galaxies for each redshift bin are needed to reconstruct the convergence
field.

There are also several theoretical uncertainties about the use of the
fractional area statistic.  One important issue is the theoretical
understanding of the one-point PDF of the smoothed convergence field.
Similar to the case of the universal halo mass function, the one-point
PDF has not been tested extensively for different cosmologies.  For
instance, the expression given by \citet{do06}, though in excellent
agreement with N-body simulations, has only been checked for one flat
$\Lambda$CDM cosmology with $\Omega_m=0.3$.  As the above discussion
indicates, the theoretical prediction for the PDF has to be accurate at
the $\sim$ 1 percent level.  It is conceivable that the universality
might break down at this precision.  However, the one-point PDF is such
a simple statistic and its derivation adds almost no extra computational
cost, once WL simulations are made. We expect our work will inspire
currently on-going or planned large WL simulations to obtain accurately
calibrated formulas for the PDF.

The baryons can cool and collapse into dense regions, which changes the
matter distribution within the virialized objects.  The effect on the
two-point statistics has been shown to be significant
\citep{mw04,zk04,jin06,rzk07}, and will therefore have to be modeled in
the interpretation of future WL surveys.  This baryonic effect would
also alter the PDF tail, and thus the fractional area statistic,
especially when small smoothing angles are used.  However, \citet{zrh07}
have shown that it is possible to take into account this uncertainty
through ``self-calibration'', i.e., constraining simultaneously the dark
energy properties and the uncertain halo profile due to baryonic
physics, e.g., halo concentration.  We hope to explore the utility of a
similar self-calibration technique for the one-point statistic, using
the halo model approach \citep{ks00} in the future.

The mass-sheet degeneracy of weak lensing is not an uncertainty for the
fractional area statistic.  Because the full PDF is measured, any offset
can be determined as the mean should be zero.  Besides, one can in
principle use the one-point PDF of the filtered tangential shear
\citep{vmb05} directly instead of the convergence PDF.

It is also important to compare the fractional area statistic with other
WL statistics.  We have found that the fractional area statistic, in
combination with CMB anisotropy measurements, may be able to reach a
cosmological sensitivity closely approaching those from using a sample
of $\sim 200,000$ shear-selected galaxy clusters \citep{us}.  The two
methods are very similar and related in practice, as the majority of the
high $S/N$ points are candidates for being real galaxy clusters
\citep{rkjs99,jv00,wk03}.  The most important gain for our approach is
that projection effects are not a contamination, since they depend on
cosmology, and therefore serve as a source of information instead.  The
fractional area statistic does not suffer from the missing or false
cluster problem, which is indeed the main motivation of this work.  The
trade-off, however, is that counting the number of clusters as a
function of redshift plays a crucial role to constrain the evolution of
dark energy, while the convergence field is a two-dimensional projection
of all the structures along the line of sight.  The tomography of the
source galaxies does help, but the broad lensing kernel entangles the
information from a wide range of redshifts; tomography is also limited
by the number of faint galaxies detected within each redshift bin (in
order to reconstruct a convergence field that is not shot-noise
dominated).

The WL power spectrum tomography \citep{sk04,zh06} and the bispectrum
tomography \citep{tj04} have been shown to be very powerful cosmological
probes and sensitive to the evolution of dark energy.  For example,
\citet{zh06} find that an LSST-like survey can deliver constraints of
$\Delta w_0\sim 0.16$ and $\Delta w_a\sim 0.36$ from lensing
alone\footnote{Note these numbers are not directly comparable to ours,
as the analysis in \citet{zh06} has a larger parameter set including,
e.g., curvature and running of the spectral index; his analysis also
utilizes the full Planck Fisher matrix instead of priors, and
marginalized over 80 photometric redshift error parameters, which we did
not do.}.  The fractional area statistic must include additional
information from higher order statistics, because the one-point PDF is
determined by all orders of cumulants \citep{bs89}.  Furthermore, in
analogy with cluster counts, the fractional area statistic is helped by
the exponential sensitivity of the tail of the PDF to the cosmological
parameters.  The scale information, imprinted in different multipole
moments of power spectrum, can be retrieved with different smoothing
scales for the fractional area statistic: the linear regime of the power
spectrum should be recovered by using various larger smoothing angular
scales and the nonlinear regime by smaller scales.  However, unlike in
Fourier space where different modes are independent, using different
smoothing scales in real space produces mass maps that are correlated
with each other.  In this work, we only utilize one particular smoothing
scale $\theta_G=1$ arcmin, which is within the nonlinear regime.  As
mentioned earlier, in the linear regime, most of the statistical
information is already contained in the two-point correlation function.
Interesting future work includes studying how to correctly take
advantage of the information from the scale-dependence of the fractional
area statistic, and its complementarity to other WL statistics.

Finally, we note that the fractional area statistic is known as the
first Minkowski functional.  In the two-dimensional case, there are two
other functionals: the length and the genus of the excursion set.  These
three Minkowski functionals, which are additive and are invariant under
translations as well as rotations, completely characterize the
morphological properties of the high convergence regions.  These
functionals have been extensively studied for CMB as useful tests of the
Gaussianity of the primordial density perturbation field [see, e.g.,
\citet{wk98}].  Since the convergence field is definitely non-Gaussian,
including the other two functionals which depend on derivatives of the
convergence field must yield additional cosmological information
\citep{lognor,gui02}.
\section{Conclusions}
\label{sec:IV}
We have shown that, in future wide field weak gravitational lensing
surveys, a simple one-point statistic -- the total fractional area of
high $S/N$ points in the convergence field -- is a promising probe of
cosmology.  It is sensitive to the total matter content of the universe
and the amplitude of the density fluctuations, which helps breaking the
intrinsic degeneracies in the growth of structure between cosmological
parameters, thus constraining the properties of dark energy.

The main conclusion of this work is that the fractional area statistic
provides constraints on cosmological parameters similar to the redshift
distribution of galaxy cluster abundance, but without suffering from the
projection effects.  Indeed, the statistic is explicitly constructed to
take advantage of projections as part of the signal.

We expect the fractional area statistic will help achieve the goal of
``precision cosmology'' and shed light on the mystery of dark energy.
\acknowledgments
The authors thank Scott Dodelson, Lam Hui, Wayne Hu, Henk Hoekstra, and
Jun Zhang for insightful discussions and comments.  This work is
supported in part by the U.S. Department of Energy under Contract
No. DE-AC02-98CH10886 and by the National Science Foundation (NSF)
through grant AST-0507161.  SW is supported by the KICP under the NSF
grant PHY-0114422.  This work is also supported in part by the
Initiatives in Science and Engineering (ISE) program at Columbia
University, and by the Pol\'anyi Program of the Hungarian National
Office for Research and Technology (NKTH).
%
%\vspace{+1\baselineskip}
\appendix
%\begin{center}
%  {\bf APPENDIX}
%\end{center}

%
In the appendices below, we give the expressions of the two-point
auto-correlation function and cross-correlation function of the
convergence fields using Limber's approximation \citep{kai92}.  We also
present a way to calculate the covariance matrix of the fractional area
of the excursion sets, assuming the true convergence field with Gaussian
smoothing can be approximated as a log-normal random field.  The
calculations below are carried out in the two-dimensional case and all
the random fields are assumed statistically isotropic.

\section{Limber's Approximation}
In this section, the two-point cross-correlation function (CCF) and
auto-correlation function (ACF) of the true convergence field are
calculated by using Limber's approximation.  Note that the
calculations below do not require the log-normal assumption.

The cross-power spectrum of the convergence field on two different
source planes at $z_i$ and $z_j$, both with Gaussian smoothing, is
related to the three-dimensional matter power spectrum $P_\delta(k)$ by
the Fourier space analogue of Limber's equation \citep{kai92}:
\be
\hat{P}_{(S)}(k_\bot)=\int^{\infty}_0d\chi\,
\frac{\cw(\chi,\chi_i)\cw(\chi,\chi_j)}{\chi^2}P_\delta\left(
\frac{k_\bot}{\chi},\chi\right)\tilde{W}^2_G(k_\bot\theta_G),
\ee
where $\chi_i\equiv\chi(z_i)$; $\cw(\chi,\chi_i)$ is given by
Eq.~(\ref{eq:wink}); $\tilde{W}_G(x)=\exp(-x^2/4)$ is the Fourier
transform of the Gaussian smoothing window function; $k_\bot$ denotes
the Fourier modes perpendicular to the line of sight, which are the only
contribution to the projected power under Limber's approximation.

The CCF is then the Fourier transform of the cross-power spectrum:
\be
\hat{C}_{(S)}(\theta)\equiv\langle\kappa_{(S)}(\bt_1,z_i)
\kappa_{(S)}(\bt_2,z_j)\rangle=\int^{\infty}_0d\chi\,
\cw(\chi,\chi_i)\cw(\chi,\chi_j)\left[\pi\int^{\infty}_0dk'_\bot\,
\frac{\Delta^2(k'_\bot,z)}{k'^{\,2}_\bot}J_0(k'_\bot\chi\theta)
\tilde{W}^2_G(k'_\bot\chi\theta_G)\right],
\label{eq:ccfln}
\ee
where we have changed variable $k'_\bot=k_\bot/\chi$.  Here
$\Delta^2(k,z)=4\pi k^3P_\delta(k,z)/(2\pi)^3$ is the dimensionless
nonlinear power spectrum, which is calculated as in~\cite{smi03}; $J_0$
denotes the Bessel function of order 0; $\theta\equiv|\bt_1-\bt_2|$ is
the angular separation of two field points.

The auto-power spectrum and the ACF of $\kappa_{(S)}$ are special cases
of the expressions above when $i=j$:
\bea
\nn
P_{(S)}(k_\bot)&=&\int^{\infty}_0d\chi\,\frac{\cw^2(\chi,\chi_s)}{\chi^2}
P_\delta\left(\frac{k_\bot}{\chi},\chi\right)
\tilde{W}^2_G(k_\bot\theta_G),\\C_{(S)}(\theta)&=&
\int^{\infty}_0d\chi\,\cw^2(\chi,\chi_s)\left[\pi\int^{\infty}_0dk'_\bot\,
\frac{\Delta^2(k'_\bot,z)}{k'^{\,2}_\bot}J_0(k'_\bot\chi\theta)
\tilde{W}^2_G(k'_\bot\chi\theta_G)\right].
\label{eq:acfln}
\eea
\section{Log-normal Random Field}
There is one specific class of non-Gaussian random fields of
theoretical interest, which can be related in some functional way to one
or several Gaussian random fields [see, e.g., \citet{cb87}; also see
\citet{bbks} and \citet{be87} for a general introduction of Gaussian
random field].  The useful feature of these non-Gaussian fields is that
when the functional form of this mapping is known, the statistical
properties are analytically determined from the underlying Gaussian
fields.

A log-normal random field is the one for which the mapping from a
Gaussian random field and the inverse are given by
\bea
\nn
\kappa&=&F_{LN}(\alpha)=|\kappa_{\rm min}|\left[\exp\left(
\sigma^{~}_{LN}\alpha-\frac{\sigma^2_{LN}}{2}\right)-1\right],\\
\alpha&=&F^{-1}_{LN}(\kappa)=\frac{1}{\sigma_{LN}}\left[\ln(1+
\frac{\kappa}{|\kappa_{\rm min}|})+\frac{\sigma^2_{LN}}{2}\right],
\label{eq:lnfun}
\eea
where $\alpha$ is a Gaussian field with zero mean and unit variance.
The minimum of the distribution $\kappa_{\rm min}$ and the variance
$\sigma^2_{LN}\equiv\ln(1+\langle\kappa^2\rangle/\kappa^2_{\rm min})$
are needed in order to specify $\kappa$.  Some relevant statistical
properties of the log-normal field $\kappa$ are listed below:
\begin{enumerate}
\item
The one-point PDF of a log-normal field $\kappa$ is obtained from the
corresponding Gaussian PDF of $\alpha$ by simply changing variable:
\be
P_1(\kappa)\,d\kappa=\frac{1}{\sqrt{2\pi\sigma^2_{LN}}}\exp\left\{
-\frac{[\ln(1+\kappa/|\kappa_{\rm min}|)+\sigma^2_{LN}/2]^2}
{2\sigma^2_{LN}}\right\}\frac{d\kappa}{|\kappa_{\rm min}|+\kappa}.
\label{eq:lnpdf}
\ee
\item
The two-point CCF of two log-normal fields $\kappa$ and $\kappa'$ is
obtained by using the joint two-point PDF of the underlying Gaussian
fields $\alpha$ and $\alpha'$:
\bea
\nn
P_2[\alpha,\alpha';\hat{C}_G(\theta)]\,d\alpha d\alpha'&=&
\frac{e^{-Q_2/2}}{\sqrt{(2\pi)^2[1-\hat{C}^2_G(\theta)]}}\,d\alpha
d\alpha',\\Q_2&=&\frac{1}{1-\hat{C}^2_G(\theta)}[\alpha^2-
2\hat{C}_G(\theta)\alpha\alpha'+\alpha'^2],
\label{eq:2ptpdf}
\eea
where $\hat{C}_G(\theta)\equiv\langle\alpha(\bt_1)\alpha'(\bt_2)\rangle$
is the CCF of $\alpha$ and $\alpha'$; $\theta$ is the angular
separation.  Carrying out the average gives the relation between these
two CCF's:
\bea
\nn
\hat{C}(\theta)&\equiv&\langle\kappa(\bt_1)\kappa'(\bt_2)\rangle=
\exp[\sigma^{~}_{LN}\sigma'_{LN}\hat{C}_G(\theta)]-1,\\
\hat{C}_G(\theta)&=&\frac{1}{\sigma^{~}_{LN}\sigma'_{LN}}
\ln[1+\hat{C}(\theta)].
\label{eq:lncc}
\eea
\item
The relation between the two-point ACF of a log-normal field and the
underlying Gaussian field is a special case of the expression above when
$\kappa'$ is just $\kappa$:
\bea
\nn
C(\theta)&=&\exp[\sigma^2_{LN}C_G(\theta)]-1,\\
C_G(\theta)&=&\frac{1}{\sigma^2_{LN}}\ln[1+C(\theta)],
\label{eq:lnac}
\eea
where $C$ and $C_G$ are the two-point ACF's of $\kappa$ and $\alpha$
respectively.  When $\theta=0$,
$C(0)=\langle\kappa^2\rangle/\kappa^2_{\rm min}$ thus
$C_G(0)=\langle\alpha^2\rangle=1$ as expected.
\end{enumerate}
\section{Statistical Properties of the Fractional Area}
In this section, we want to calculate the covariance matrix in
Eq.~(\ref{eq:fisher}) by assuming that the convergence field with
Gaussian smoothing can be approximately described by a log-normal
random field $\kappa_{(S)}(\alpha)=F_{LN}(\alpha)$.  To specify the
function $F_{LN}$, the parameter $\kappa_{\rm min}$ is given by
Eq.~(\ref{eq:kmin}) and the other parameter $\sigma^2_{LN}$ can be
calculated using Eq.~(\ref{eq:acfln}):
\bea
\nn
\kappa_{\rm min}&=&-\int^{\infty}_0d\chi\,\cw(\chi,\chi_s),\\
\langle\kappa^2_{(S)}\rangle=C_{(S)}(0)&=&\int^{\infty}_0d\chi\,
\cw^2(\chi,\chi_s)\left[\pi\int^{\infty}_0dk'_\bot\,
\frac{\Delta^2(k'_\bot,z)}{k'^{\,2}_\bot}
\tilde{W}^2_G(k'_\bot\chi\theta_G)\right],
\label{eq:varkpa}
\eea
as $J_0(0)=1$.  Note that they both depend on the source redshift
$z_s$.  The CCF of the underlying Gaussian field (similarly for ACF)
is obtained by using Eq.~(\ref{eq:lncc}), where $\hat{C}(\theta)$ is
now $\hat{C}_{(S)}(\theta)$ as given in Eq.~(\ref{eq:ccfln}).

The mean fractional area of the excursion set $E_\nu$ under the
log-normal assumption is given by
\be
\langle\cf(\nu,z_s)\rangle=\frac{1}{2}\mathrm{erfc}
\left(\frac{x_\nu}{\sqrt{2}}\right),
\label{eq:a1ln}
\ee
where erfc is the complementary error function;
$x_\nu\equiv F^{-1}_{LN}(\nu\sigma_{(N)})$ is the effective
threshold for $\alpha$.

The covariance between the fractional area with threshold of $\mu$ at
redshift $z_i$ and the one with threshold of $\nu$ at $z_j$ is given by
\be
\mathrm{Cov}[\cf(\mu,z_i),\cf(\nu,z_j)]=
\frac{1}{A^2_{\rm tot}}\left(\langle A_\mu A_\nu\rangle-
\langle A_\mu\rangle\langle A_\nu\rangle\right),
\ee
where the statistical average is taken by using the joint two-point
PDF as given in Eq.~(\ref{eq:2ptpdf}).  Because the two-point PDF only
depends on the angular separation $\theta$ between two different lines
of sight, one spatial integral can be eliminated and gives
\bea
\nn
\mathrm{Cov}[\cf(\mu,z_i),\cf(\nu,z_j)]&=&
\frac{1}{A_{\rm tot}}\int^{\theta_{\rm max}}_02\pi\theta\,d\theta
\int_{\mu\sigma_{(N)}}^{\infty}d\kappa_{1(S)}(z_i)
\int_{\nu\sigma_{(N)}}^{\infty}d\kappa_{2(S)}(z_j)\,
P_2[\kappa_{1(S)}(z_i),\kappa_{2(S)}(z_j);\theta]
-\langle\cf(\mu,z_i)\rangle\langle\cf(\nu,z_j)\rangle\\
&=&\frac{1}{A_{\rm tot}}\int^{\theta_{\rm max}}_02\pi\theta\,d\theta
\left\{\int_{x_\mu}^{\infty}d\alpha_i\int_{x_\nu}^{\infty}d\alpha_j\,
P_2[\alpha_i,\alpha_j;\hat{C}_G(\theta)]
-\langle\cf(\mu,z_i)\rangle\langle\cf(\nu,z_j)\rangle\right\},
\label{eq:tmpcov}
\eea
where we have moved the $\langle\cf\rangle\langle\cf\rangle$ term inside
the integral over $\theta$, since it has no dependence on $\theta$;
$x_\mu$ and $x_\nu$ are the effective thresholds of $\alpha_i$ and
$\alpha_j$ respectively as defined in Eq.~(\ref{eq:a1ln}).

Let us denote
\be
\langle\cf_2[\mu,z_i;\nu,z_j;\hat{C}_G(\theta)]\rangle=
\int^{\infty}_{x_\mu}d\alpha_i\int^{\infty}_{x_\nu}d\alpha_j\,
P_2[\alpha_i,\alpha_j;\hat{C}_G(\theta)].
\label{eq:a2ln}
\ee
The special case of this expression with $\mu=\nu$ and $i=j$, had been
extensively studied \citep{kai84,pw84,js86,kas91} to approximate the
correlation function of ``biased regions'', e.g., galaxy clusters.
Evaluation of this integral hinges on the equality \citep{hgw86,wk98}
\be
\frac{\partial\langle\cf_2(\hat{C}_G)\rangle}{\partial\hat{C}_G}=
P_2[x_\mu,x_\nu;\hat{C}_G],
\ee
which can be proven by Fourier transforming the integrand $P_2$, then
integrating over $\alpha$'s and Fourier transforming back.  This
first-order ordinary differential equation has a boundary condition
that, when $\hat{C}_G=0$, the two-point PDF reduces to the product of
two independent one-point PDF.  Then the double integral can be
separated and gives
\be
\langle\cf_2(\mu,z_i;\nu,z_j;0)\rangle=
\langle\cf(\mu,z_i)\rangle\langle\cf(\nu,z_j)\rangle.
\ee
Thus the term in the curly braces of Eq.~(\ref{eq:tmpcov}) is simply
\be
\langle\cf_2[\mu,\nu;z_i,z_j;\hat{C}_G(\theta)]\rangle-
\langle\cf_2(\mu,\nu;z_i,z_j;0)\rangle=
\int^{\hat{C}_G(\theta)}_0dC'\,P_2(x_\mu,x_\nu;C').
\ee

Substituting the above expression into Eq.~(\ref{eq:tmpcov}), and
changing the order of integration by assuming $\hat{C}_G(\theta)$ is
monotonic\footnote{Here we have assumed that the correlation function
stays positive and monotonically decreases to zero as the scale becomes
large.  However, the actual matter correlation function has to turn
negative at a particular scale, before approaching zero.  This is
because if matter is clustered on small scales, then they have to be
``anti-clustered'' on large scales to conserve the total amount of mass.
This contribution has an opposite sign, which makes the scale-dependence
of the variance of the fractional area steeper than Poisson-scaling,
i.e., the variance goes to zero faster than $\theta^{-2}$.}, we get:
\be
\mathrm{Cov}[\cf(\mu,z_i),\cf(\nu,z_j)]=\frac{1}{A_{\rm tot}}
\int^{\hat{C}_G(0)}_0dC'\,\frac{\hat{\theta}^2(C')}{2\sqrt{1-C'^2}}
\exp\left[-\frac{x^2_\mu-2C'x_\mu x_\nu+x^2_\nu}{2(1-C'^2)}\right].
\label{eq:cov}
\ee
Here $\hat{\theta}(C')$ is the inverse of the CCF: $\hat{C}_G(\theta)$
between $\alpha$'s as given in Eq.~(\ref{eq:lncc}).  The covariance
reduces to the variance when $\mu=\nu$ and $z_i=z_j$:
\be
\mathrm{Var}[\cf(\nu,z_s)]=\frac{1}{A_{\rm tot}}\int^{C_G(0)}_0dC'\,
\frac{\theta^2(C')}{2\sqrt{1-C'^2}}\exp\left(-\frac{x^2_\nu}{1+C'}\right),
\ee
where $\theta(C')$ is the inverse of the ACF: $C_G(\theta)$ as given in
Eq.~(\ref{eq:lnac}).  These integrals are finite as long as the CCF/ACF
falls off faster than $\theta^{-2}$ at large distances.

We have also assumed $\hat{C}_G(\theta_{\rm max})\sim 0$, i.e., the size
of the field $\theta_{\rm max}$ is large enough compared with the
correlation length of the convergence field [see Eq.~(\ref{eq:corrlg})].
The correction term of Eq.~(\ref{eq:cov}) is
\be
\Delta(\mathrm{Cov})=\int^{\hat{C}_G(\theta_{\rm max})}_0dC'\,
\frac{1-\hat{\theta}^2(C')/\theta^2_{\rm max}}{2\pi\sqrt{1-C'^2}}
\exp\left[-\frac{x^2_\mu-2C'x_\mu x_\nu+x^2_\nu}{2(1-C'^2)}\right].
\label{eq:covd}
\ee
\section{Random Noise Field}
In the simplest case where the noise due to intrinsic ellipticity and
signal are independent of each other, the one-point PDF of the noisy
convergence field $K=\kappa_{(S)}+\kappa_{(N)}$ is just a convolution of
the log-normal and Gaussian distribution.  So the mean fractional area
of the excursion set $E_\nu$ with $K>\nu\sigma_{(N)}$ is
\be
\langle\cf(\nu,z_s)\rangle=\int_{-\infty}^{+\infty}
\frac{d\kappa_{(N)}}{\sqrt{2\pi\sigma^2_{(N)}}}\exp\left[-
\frac{(\kappa_{(N)}-\kappa_{\rm bias})^2}{2\sigma^2_{(N)}}\right]\,
\frac{1}{2}\mathrm{erfc}\left(\frac{x_\nu}{\sqrt{2}}\right),
\ee
where the effective threshold
$x_\nu\equiv F^{-1}_{LN}(\nu\sigma_{(N)}-\kappa_{(N)})$ now depends on
the integral variable $\kappa_{(N)}$.  We define
$F^{-1}_{LN}(\kappa)=-\infty$ when $\kappa\leq\kappa_{\rm min}$.  We
also allow a nonzero mean of the noise field $\kappa_{\rm bias}$ as in
the self-calibration case.

Similarly for the covariance, the integral in Eq.~(\ref{eq:a2ln}) will
be taken over two more variables due to the noise field.  The
expression involves a five-dimensional integral which, unfortunately,
cannot be simplified as what is done in the noise-free case and is
very messy.  Thus it is useful to get an upper bound of the covariance
for this general case as in \citet{wk98}.  Note that in the limit when
$x_\mu$ or $x_\nu$ goes to $-\infty$, $\langle\cf_2\rangle$ is simply
$\langle\cf\rangle$.  Since the two-point PDF $P_2$ is always
non-negative, we get
$\langle\cf_2\rangle\leq\min[\langle\cf(\mu,z_i)\rangle,\langle\cf(\nu,z_j)\rangle]$.
By using this inequality, the integrand of Eq.~(\ref{eq:tmpcov}) then
has no $\theta$ dependence.  Therefore,
\be
\mathrm{Cov}[\cf(\mu,z_i),\cf(\nu,z_j)]\leq
\frac{\pi\theta^2_c}{A_{\rm tot}}
\langle\cf\rangle_<(1-\langle\cf\rangle_>),
\label{eq:bound}
\ee
where the subscript ``$<$'' (or ``$>$'') denotes the smaller (or larger)
one between two $\langle\cf\rangle$'s.  The correlation length
$\theta_c$ is the scale beyond which the correlation can be neglected
for physical reasons and is defined as \citep{be87}
\be
\theta^2_c\equiv\frac{\langle K^2\rangle}{\langle(\nabla K)^2\rangle}=
\frac{\langle\kappa^2_{(S)}\rangle+\langle\kappa^2_{(N)}\rangle}
{\langle(\nabla\kappa_{(S)})^2\rangle+\langle(\nabla\kappa_{(N)})^2\rangle},
\label{eq:corrlg}
\ee
where the variances of the derivative fields are given by
\bea
\nn
\langle(\nabla\kappa_{(S)})^2\rangle&=&-\nabla^2C_{(S)}(0)=\int^{\infty}_0
d\chi\,\chi^2\cw^2(\chi,\chi_s)\left[\pi\int^{\infty}_0dk'_\bot\,
\Delta^2(k'_\bot,z)\tilde{W}^2_G(k'_\bot\chi\theta_G)\right],\\
\langle(\nabla\kappa_{(N)})^2\rangle&=&-\nabla^2C_{(N)}(0)=
\frac{2\sigma^2_{(N)}}{\theta^2_G},
\eea
as $\nabla^2J_0(\theta)=-J_0(\theta)$.  We again assume
$\theta_{\rm max}\gg\theta_c$ here.

\end{document}